\documentclass[aps,pra,twocolumn,superscriptaddress,longbibliography]{revtex4-2}
\usepackage{graphicx}
\usepackage{latexsym}
\usepackage{amssymb}
\usepackage{amsmath}
\usepackage{amsfonts}
\usepackage{upgreek}
\usepackage{float}
\usepackage{bm}
\usepackage{multirow}
\usepackage{color}
\usepackage[T1]{fontenc}
\usepackage{hyperref}
\usepackage{subfigure}
\usepackage{booktabs,tabularx}

\hypersetup{
colorlinks = true,
linkcolor = [rgb]{0.70,0.13,0.13},
citecolor = [rgb]{0.13,0.55,0.13},
urlcolor  = [rgb]{0.25, 0.41, 0.88}}

\begin{document}

\title{Deconfined quantum criticality of frustrated hard-core dipolar bosons}

\author{Ya-Nan Wang}
\affiliation{College of Physics, Nanjing University of Aeronautics and Astronautics, Nanjing, 211106, China}
\affiliation{Key Laboratory of Aerospace Information Materials and Physics (Nanjing University of Aeronautics and Astronautics), MIIT, Nanjing 211106, China}

\author{Wen-Long You}
\affiliation{College of Physics, Nanjing University of Aeronautics and Astronautics, Nanjing, 211106, China}
\affiliation{Key Laboratory of Aerospace Information Materials and Physics (Nanjing University of Aeronautics and Astronautics), MIIT, Nanjing 211106, China}

\author{Wen-Yi Zhang}
\affiliation{College of Physics, Nanjing University of Aeronautics and Astronautics, Nanjing, 211106, China}
\affiliation{Key Laboratory of Aerospace Information Materials and Physics (Nanjing University of Aeronautics and Astronautics), MIIT, Nanjing 211106, China}

\author{Su-Peng Kou}
\affiliation{School of Physics and Astronomy, Beijing Normal University, Beijing 100875, China}

\author{Gaoyong Sun}
\thanks{Corresponding author: gysun@nuaa.edu.cn}
\affiliation{College of Physics, Nanjing University of Aeronautics and Astronautics, Nanjing, 211106, China}
\affiliation{Key Laboratory of Aerospace Information Materials and Physics (Nanjing University of Aeronautics and Astronautics), MIIT, Nanjing 211106, China}

\begin{abstract}
Deconfined quantum critical points (DQCPs) are proposed as unconventional second-order phase transitions beyond the Landau-Ginzburg-Wilson paradigm.
The nature and experimental realizations of DQCPs are crucial issues of importance.
We illustrate the potential for DQCPs between the valence bond solid state and the antiferromagnetic phase to arise in optical lattices containing frustrated dipolar bosons subject to hard-core constraints. 
The emergence of DQCPs is comprehended through the fusion of two Berezinskii-Kosterlitz-Thouless (BKT) transitions.
The DQCPs and the BKTs are confirmed by the scaling of ground-state fidelity susceptibilities in finite systems and the analysis of order parameters obtained from infinite systems.
The numerical analysis reveals varying critical exponents of the correlation length in DQCPs and the logarithmic scaling in BKTs, respectively.
This work offers a promising platform for realizing DQCPs and provides valuable insights into their nature within the framework of topological phase transitions.

\end{abstract}

\maketitle

\section{Introduction}
The quantum phase transition is one of the fundamental concepts in realms of statistical physics and condensed matter physics \cite{sachdev1999quantum}. 
Continuous phase transitions are effectively characterized by the Landau-Ginzburg-Wilson (LGW) paradigm through the utilization of local order parameters based on the concept of spontaneous symmetry breaking.
Berezinskii–Kosterlitz–Thouless (BKT) transitions \cite{berezinskii1972destruction,kosterlitz1973ordering} and deconfined quantum critical points (DQCPs) \cite{senthil2004deconfined,senthil2004quantum} stand as two renowned examples that extend beyond the framework of the LGW theory.
BKT transitions are infinite-order phase transitions that occur between quasi-ordered phases and disordered phases \cite{berezinskii1972destruction,kosterlitz1973ordering}, while DQCPs represent second-order phase transitions between two ordered phases \cite{senthil2004deconfined,senthil2004quantum}.
Over last decades, DQCPs have garnered significant attention, particularly in exploring the nature of phase transitions (first or second order) \cite{senthil2004deconfined,senthil2004quantum,sandvik2007evidence}, quantum criticalities \cite{shao2016quantum}, duality phenomena \cite{qin2017duality,janssen2017critical,zhang2023exactly,wang2017deconfined}, and emergent symmetries \cite{wang2017deconfined,nahum2015emergent,huang2019emergent}.
In previous studies, BKT transitions and DQCPs were investigated separately. Here, we demonstrate that DQCPs can arise from the fusion of two BKT transitions, offering essential insights into comprehending both BKT transitions and DQCPs.

\begin{figure}[t]
\includegraphics[width=8.3cm]{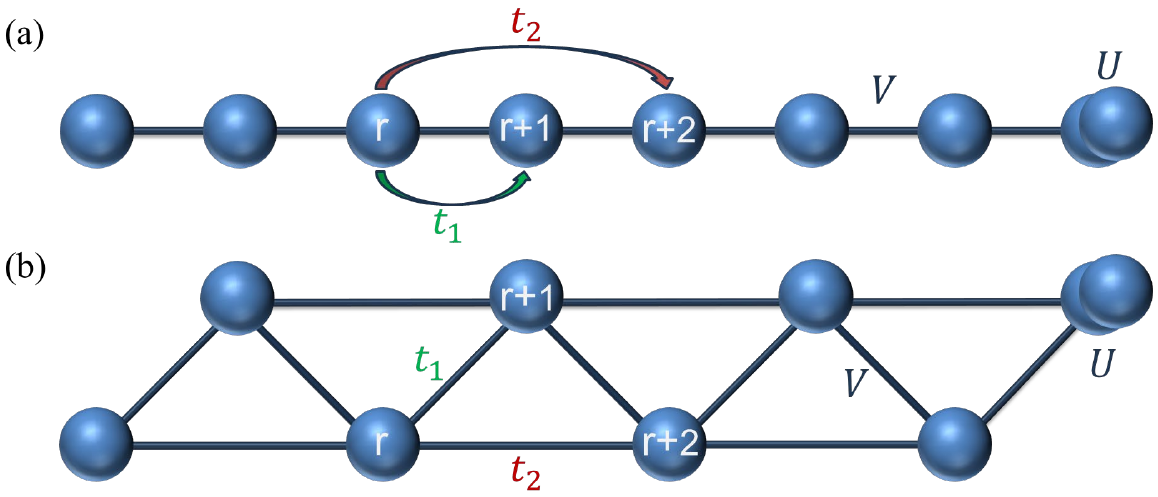} 
\centering
\caption{ Geometric structures of the frustrated dipolar Bose-Hubbard model. 
(a) The one-dimensional chain involves the nearest-neighbor hopping $t_1$ and the next-nearest-neighbor hopping $t_2$, on-site interaction $U$ and nearest-neighbor interaction $V$. 
(b) The zig-zag chain corresponding to (a). }
\label{modelfig}
\end{figure}

Despite the wide range of theoretical explorations on DQCPs \cite{senthil2023deconfined}, there are relatively few experimental realizations \cite{cui2023proximate,song2024unconventional}.
Ultracold bosons, serving as an ideal quantum simulator, provide a versatile platform for engineering many-body physics \cite{lahaye2009physics,goral2002quantum}.
Taking into account the interplay of interactions and lattice geometry, ultracold bosons unveil numerous rich novel quantum phenomena \cite{greiner2002quantum,bloch2008many,jaksch1998cold,lewenstein2007ultracold,dutta2015non}.
The zig-zag lattice, being the simplest form of a triangular lattice, embodies distinctive geometric frustration and topological structure, serves as an ideal candidate for exploring quantum phases \cite{Greschner2013ultracold,sun2014topological,dhar2013hard,anisimovas2016semisynthetic,greschner2019interacting}.
For example, hard-core bosons arranged in a triangular lattice configuration can induce a supersolid phase, distinguished by the coexistence of solid and superfluid characteristics \cite{Supersolid_Hard-Core_Bosons_Wessel_2005,Supersolid_Phase_Boninsegni_2005,Persistent_Supersolid_Heidarian_2005,Supersolid_Order_from_Disorder_Melko_2005}. 
Recently, there has been significant focus on proposed experimental explorations of DQCPs within the domain of programmable quantum simulators, involving systems such as arrays of Rydberg quantum simulators \cite{lee2023landau}, trapped ions quantum simulators \cite{romen2024deconfined}, and state-dependent optical lattices with ultracold bosons \cite{baldelli2024frustrated}.

In this paper, we instead explore a frustrated dipolar Bose-Hubbard (FDBH) model in one dimension under the hard-core limit, where nearest-neighbor interactions stem from dipolar interactions \cite{su2023dipolar}, which are achievable within an atomic mixture \cite{baldelli2024frustrated} as well.
We show that the FDBH model hosts DQCPs, aligning with recent theoretical proposals \cite{huang2019emergent,Jiang2019,Roberts2019,luo2019intrinsic,sun2019fidelity} that describe a continuous quantum phase transition from an antiferromagnetic (AFM) state to a valence bond solid (VBS) state.
We employ the density matrix renormalization group (DMRG) technique \cite{white1992density,white1993density,verstraete2008matrix,schollwock2011density,orus2014practical} and the infinite time evolving block decimation (iTEBD) method \cite{vidal2007classical} to investigate quantum phase transitions of FDBH model by varying nearest-neighbor interactions.
Our study reveals the presence of three distinct phases, delineated by two BKT transitions and one DQCP through the fidelity susceptibility.
Notably, our numerical simulations confirm that DQCPs emerge from the fusion of two BKT transitions \cite{lee2023landau,romen2024deconfined,baldelli2024frustrated}, providing an essential insight into the understanding of both BKT transitions and DQCPs.

\section{Model}
We consider interacting dipolar bosons within a one-dimensional lattice accounting for both nearest-neighbor and next-nearest-neighbor hopping terms, 
as illustrated in Fig.\ref{modelfig}(a). The corresponding Hamiltonian of the FDBH model is described by
\begin{align}
H &= \sum_r \left( t_1 b_r^{\dagger}b_{r+1} + t_2 b_r^{\dagger}b_{r+2} + \text{h.c.} \right) \nonumber \\ 
   & \hspace{2.5cm} +  V n_r n_{r+1}+\frac{U}{2}n_r(n_r-1).
\label{eq:Ham}
\end{align}
In this Hamiltonian, $b_r$ ($b_r^{\dagger}$) denote the bosonic annihilation (creation) operators at the $r$th site.
The expression $n_r=b_r^{\dagger}b_r$ represents the particle number operator, with $t_1 \geq 0$ and $t_2 \geq 0$ denoting the amplitudes for nearest-neighbor and next-nearest-neighbor hopping, respectively.
The positive hopping amplitudes are employed to introduce frustrations \cite{sato2011competing}, which can be achieved using the Floquet protocol \cite{eckardt2017colloquium,sun2020optimal,kwan2024realization,impertro2024realization}. $V \geq 0$ and $U \geq 0$ signify the strengths of the nearest-neighbor and on-site interactions.
Note that the Hamiltonian in Eq.(\ref{eq:Ham}) can be realized using a synthetic dimension \cite{cabedo2020effective,barbiero2023frustrated}, where the interspecies interactions serve as dipolar interactions \cite{baldelli2024frustrated}, offering an alternative approach for realizing DQCPs. 

\begin{figure}[t]
\includegraphics[width=8.7cm]{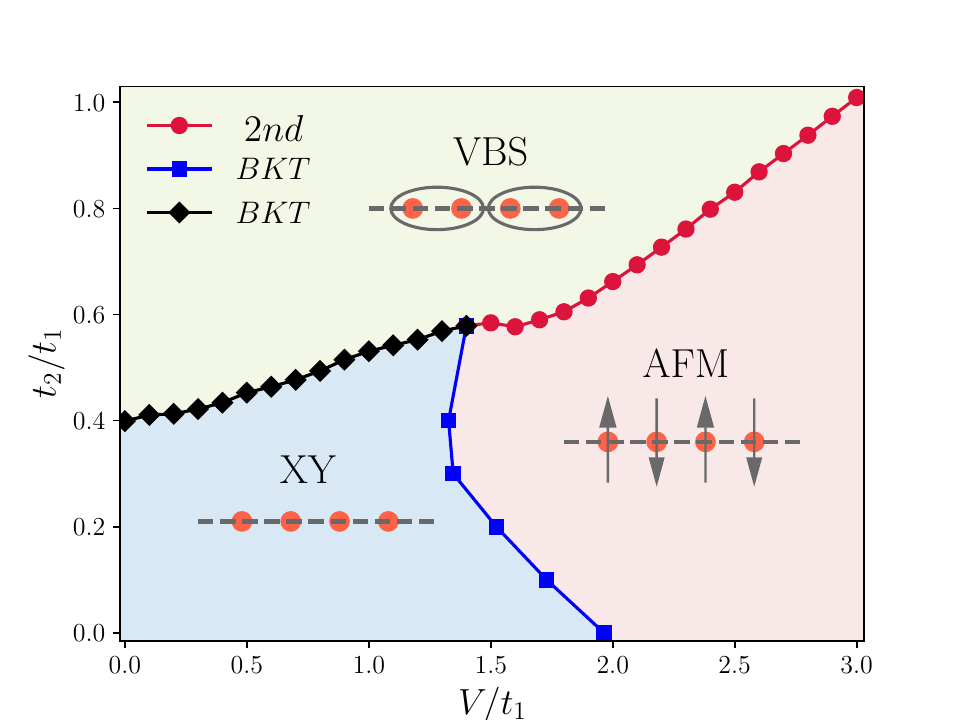} 
\centering
\caption{The phase diagram with respect to $V/t_1$ and $t_2/t_1$ of the one-dimensional FDBH model in the hard-core limit includes the VBS phase, the XY phase and the AFM phase. 
The red circles denote a second-order (2nd) phase transition (DQCP) from VBS to AFM, while the black diamonds and blue squares represent the BKT transitions from XY to VBS and XY to AFM, respectively. }
\label{phasefig}
\end{figure}

The frustrated one-dimensional chain is equivalent to the zigzag chain [cf. Fig. \ref{modelfig}(b)], which can be formed by the incoherent superposition of an optical triangular lattice \cite{Greschner2013ultracold}.
Ultracold bosons offer a wealth of physics arising from the interplay between frustrations induced by lattice geometry and interactions \cite{Greschner2013ultracold}. 
For instance, it is demonstrated that ultracold bosons subject to a three-body constraint exhibit chiral superfluid and Haldane insulator phases within zig-zag optical lattices \cite{Greschner2013ultracold}.
We note that although long-range dipolar interactions decay as $1/|r-r^{\prime}|^3$, the properties of hard-core dipoles in optical lattices at half-filling can be effectively understood in terms of nearest-neighbor interactions \cite{capogrosso2010quantum,deng2013polar,dutta2015non}. Therefore, we focus solely on the nearest-neighbor interaction term.

In the hard-core limit ($U \rightarrow \infty$), the FDBH model in Eq.(\ref{eq:Ham}) becomes
\begin{align}
H &= \sum_r \left( t_1 b_r^{\dagger}b_{r+1} + t_2 b_r^{\dagger}b_{r+2} + \text{h.c.} \right) + V n_{r} n_{r+1},
\label{eq:hardcoreHam}
\end{align}
with a constraint $n_{r}=\{ 0,1 \}$.
In the following, we will investigate the physics underlying the hard-core FDBH model, as illustrated in Eq.(\ref{eq:hardcoreHam}), wherein the total number of sites in the chain is assumed to be divisible by four in order to comprehensively account for the translation symmetry-breaking phase.

\begin{figure}[t]
\includegraphics[width=8.6cm]{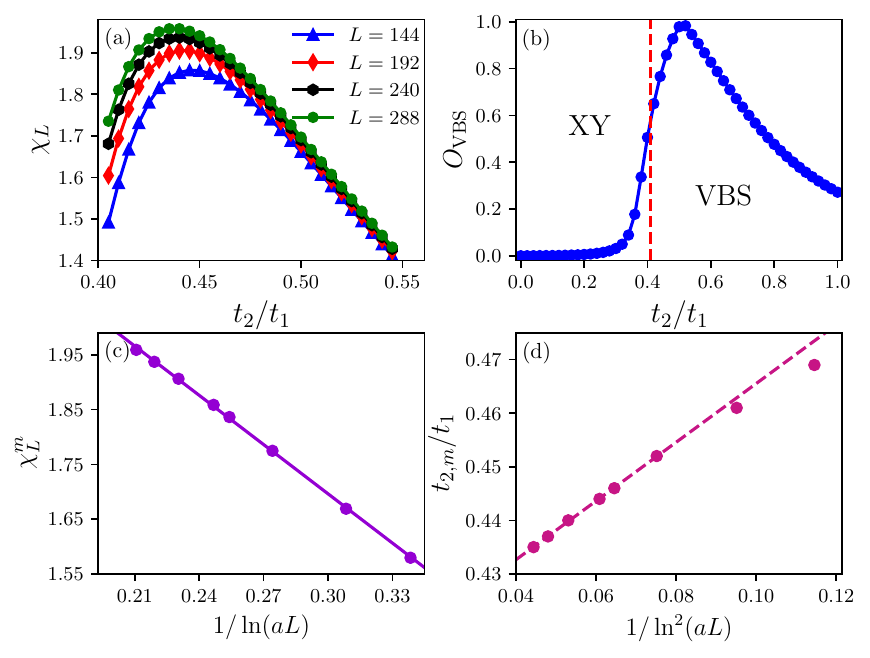} 
\centering
\caption{Phase transition between the XY phase and the VBS phase in hard-core FDBH chain at $V/t_1=0.1$ in OBCs. 
(a) Fidelity susceptibility $\chi_{L}$ is examined with respect to $t_2/t_1$, in which the peak values and positions of $\chi_{L}$ gradually increase and shift to the left, respectively, as the system size increases. 
(b) The VBS order parameter $O_{\text{VBS}}$ plotted against $t_2/t_1$ obtained from the iTEBD method. The red dashed line indicates the critical point $t_{2c}/t_1$ in the thermodynamic limit from the finite scaling of the fidelity susceptibility.
(c) The finite-size scaling of the peak values of $\chi_{L}^{m}$ with respect to the system sizes $1/\ln{(aL)}$ confirms the BKT transition, with $a$ representing the cutoff constant.
(d) The scaling of the peak positions $t_{2,m}/t_{1}$ as a function of the system sizes $1/\ln^2{(aL)}$, where the critical value obtained through the extrapolation is $t_{2c}/t_{1}=0.4107$, further confirming the presence of the BKT transition.}
\label{V0.1fig}
\end{figure}

\section{Berezinskii-Kosterlitz-Thouless transitions} 
Subject to the hard-core constraints ($U \rightarrow \infty$), the FDBH model described in Eq.(\ref{eq:hardcoreHam}) can be effectively transformed into a spin model, as presented below:
\begin{align}
H &= \sum_{r}\frac{t_1}{2}(\sigma_r^x\sigma_{r+1}^x+\sigma_r^y\sigma_{r+1}^y)+\frac{t_2}{2}(\sigma_r^x\sigma_{r+2}^x+\sigma_r^y\sigma_{r+2}^y) \nonumber \\
    & \hspace{3cm}  + \frac{V}{4}(1-\sigma_r^z)(1-\sigma_{r+1}^z),
\label{H_spin}
\end{align}
by utilizing the transformations $\sigma^{x}_{r} = b^{\dagger}_{r} + b_{r}, ~ ~ \sigma^{y}_{r} = i(b^{\dagger}_{r} - b_{r}), ~ ~ \sigma^{z}_{r} = 1 - 2n_{r}$, with $(\sigma^{x}_{r}, \sigma^{y}_{r}, \sigma^{z}_{r})$ being Pauli matrices. Here, the particle occupation numbers are represented by spins. 

\begin{figure}[tb]
\includegraphics[width=8.3cm]{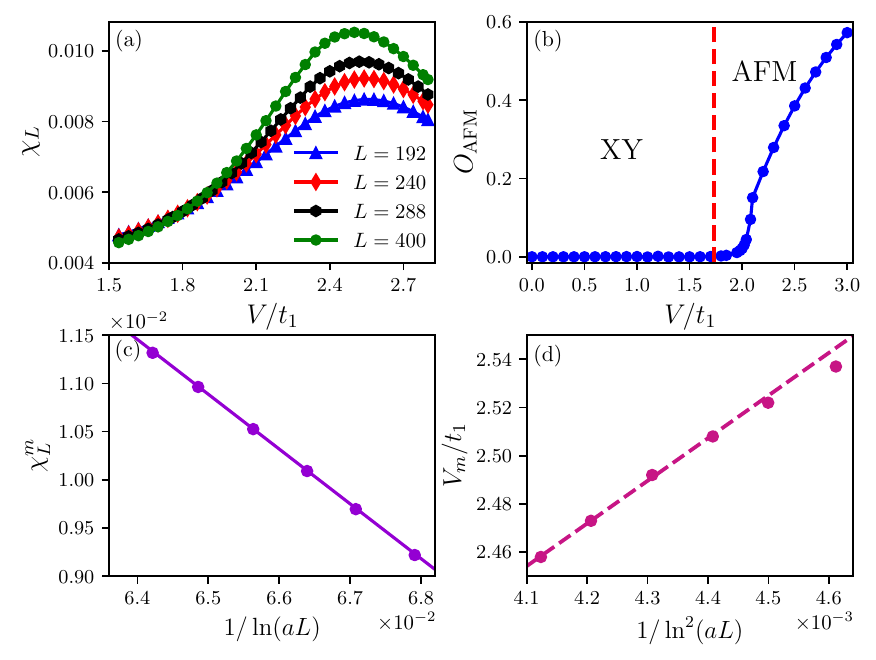} 
\centering
\caption{Phase transition between the XY phase and the AFM phase in hard-core FDBH chain at $t_{2}/t_1=0.1$. 
(a) The fidelity susceptibility $\chi_{L}$ in OBCs varies with $V/t_1$, exhibiting an increase in peak values and leftward shifts in peak positions with increasing system size.
(b) The AFM order parameter $O_{\text{AFM}}$ plotted against $V/t_1$ obtained from the iTEBD method. The red dashed line indicates the critical point $V_{c}/t_1$ in the thermodynamic limit from the finite scaling of the fidelity susceptibility.
(c) The finite-size scaling of the peak values of $\chi_{L}^{m}$ with respect to the system sizes $1/\ln{(aL)}$ confirms the BKT transition, with $a$ representing the cutoff constant.
(d) The scaling of the peak positions $V_{m}/t_{1}$ as a function of the system sizes $1/\ln^2{(aL)}$, where the critical value obtained through the extrapolation is $V_{c}/V_{1}=1.729$, further confirming the presence of the BKT transition.}
\label{tx0.1fig}
\end{figure}

When $V=0$, the system shown in Eq.(\ref{H_spin}) become the XY model incorporating both the nearest-neighbor and next-nearest-neighbor hopping terms. 
In particular, when $t_2 = 0$, the system's ground state is characterized by the gapless XY (superfluidity (SF) in bosonic language) phase (see Appendix \ref{AppA} for details). 
Upon increasing $t_2$, the system undergoes a phase transition from the XY phase to the VBS (bond-ordered-wave (BOW) in bosonic language) phase \cite{baldelli2024frustrated,mudry2019quantum,haldane1982spontaneous,okamoto1992fluid,nomura1994critical}.
To verify if the transition between the XY phase and the VBS phase persists in the presence of interaction $V$, we calculate the ground-state phase diagram of the FDBH model described in Eq.(\ref{eq:hardcoreHam}) varying parameters $t_2/t_1$ and $V/t_1$ in terms of DMRG method. During DMRG simulations, we keep 500 states and enforce $n_{\text{max}}=1$ per site. 
The full phase diagram is shown in Fig.\ref{phasefig}, which is obtained from the fidelity susceptibility \cite{you2007fidelity,Quantum_Critical_Scaling2007,gu2010fidelity,sun2017fidelity,zhu2018fidelity,chen2008intrinsic,damski2013fidelity,luo2018fidelity}, 
\begin{equation}
    \chi_L=\frac{1}{L}\lim\limits_{\delta\lambda_l \to 0} \frac{-2 \ln F(\lambda_l,\lambda_l+\delta\lambda_l)} {(\delta\lambda_l)^2}.
\end{equation}
Here, $F(\lambda_l,\lambda_l+\delta\lambda_l)=|\langle\Psi_0(\lambda_l)|\Psi_0(\lambda_l+\delta \lambda_l)\rangle|$ is the fidelity for $l=1,2$, with the control parameters defined as $\lambda_1= t_2/t_1$ and $\lambda_2=V/t_1$.
The numerical results support the argument that the phase transition persists even under weak interaction.
In Fig.\ref{V0.1fig}(a), we depict the scaling behavior of fidelity susceptibility at an interaction strength of $V/t_{1}=0.1$ by varying the next-nearest-neighbor hopping $t_2/t_1$. 
It is observed that the fidelity susceptibility increases gradually as the system's size increases. 
Furthermore, both the maximum fidelity susceptibility $\chi_L^m$ and the peak positions $\lambda_m$ themselves are demonstrated to be in excellent agreement with the scaling laws [c.f. Fig.\ref{V0.1fig}(c) and (d)] proposed for BKT transitions \cite{sun2015fidelity}, as described by:
\begin{align}
\chi_L^m \simeq \chi_{\infty} - \frac{1}{\ln(aL)}, \label{FS} \\
\lambda_m \simeq \lambda_{c} + \frac{1}{\ln^2(aL)}, \label{FS2}
\end{align}
where $\chi_{\infty}$ represents the fidelity susceptibility in the thermodynamic limit, $\lambda_c$ is the critical point, and $a$ is a positive nonuniversal constant serving as a cutoff.
The transition from the XY phase to the VBS phase is further validated through level crossing (see Appendix \ref{AppB} for details) and the order parameter [c.f. Fig.\ref{V0.1fig}(b)] characterizing the VBS phase under open boundary conditions (OBCs), as expressed in:
\begin{align}
O_{\text{VBS}} &= \frac{1}{3} \left( |\langle \bm{\sigma}_{r} \cdot \bm{\sigma}_{r+1}\rangle| - |\langle \bm{\sigma}_{r+1} \cdot \bm{\sigma}_{r+2}\rangle| \right ), \label{OP:VBS}
\end{align}
where $\bm{\sigma}_{r}=(\sigma^{x}_{r}, \sigma^{y}_{r}, \sigma^{z}_{r})$.

\begin{figure}[tb]
\includegraphics[width=9.1cm]{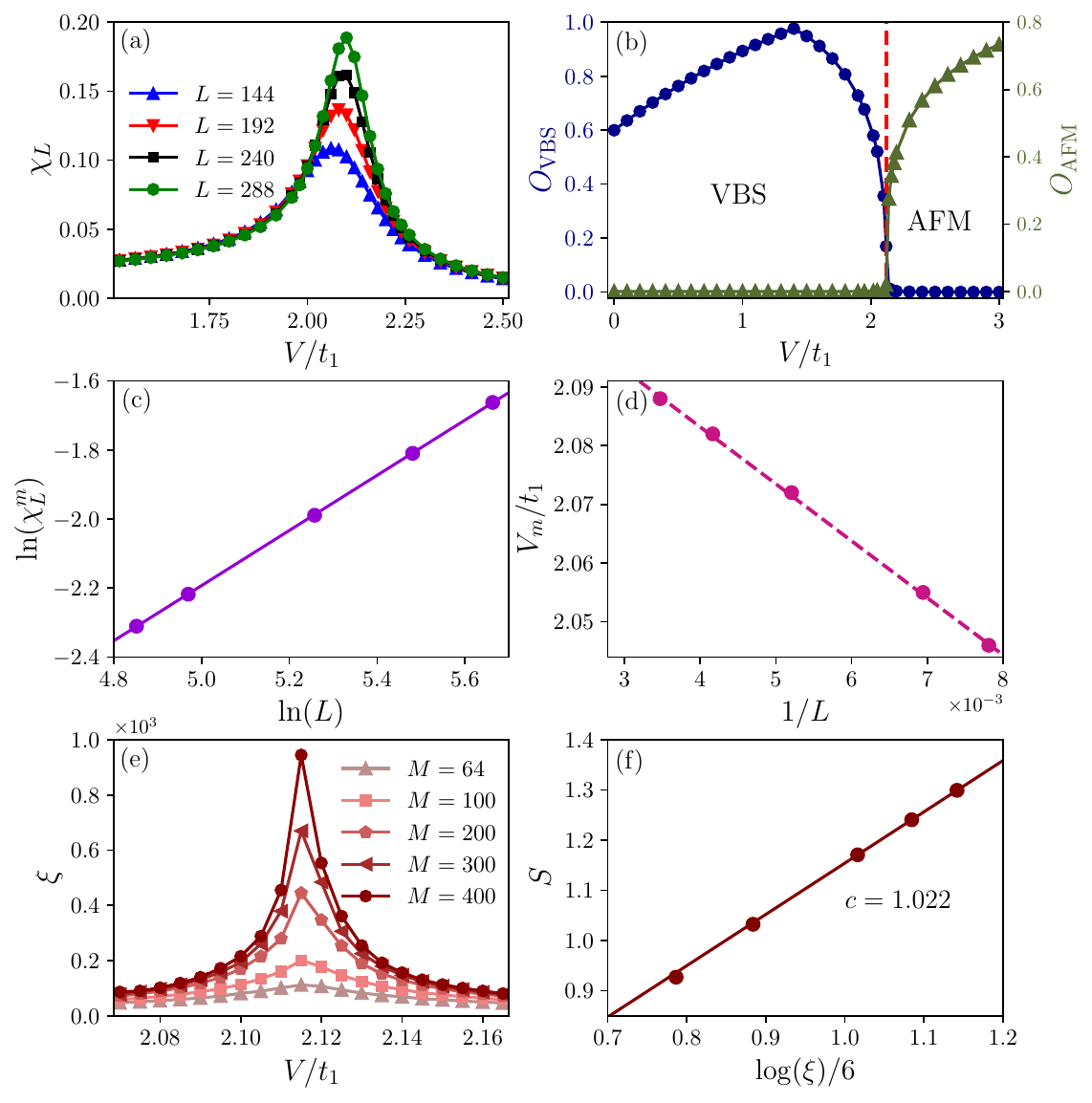} 
\centering
\caption{Phase transition between the VBS phase and the AFM phase in hard-core FDBH chain at $t_{2}/t_1=0.7$. 
(a) The fidelity susceptibility $\chi_{L}$ varies with $V$ in OBCs. The peak position of $\chi_{L}$ shifts to the right as the system size increases.
(b) The VBS and AFM order parameters $O_{\text{VBS}}$ and $O_{\text{AFM}}$ plotted against $V/t_1$ obtained from the iTEBD method. The red dashed line indicates the critical point $V_{c}/t_1$ in the thermodynamic limit from the finite scaling of the fidelity susceptibility.
(c) The finite-size scaling of the peak values of $\ln(\chi_{L}^{m})$ with respect to the system sizes $\ln(L)$ confirms the 2nd transition with the correlation length critical exponent $\nu=1.1112$.
(d) The scaling of the peak positions $V_{m}/t_{1}$ as a function of the system sizes $1/L$, where the critical value obtained through the extrapolation is $V_{c}/V_{1}=2.122$, further confirming the presence of the 2nd transition.
(e) Correlation lengths as a function of $V/t_{1}$ for different numbers of states $M=64, 100, 200, 300, 400$, respectively.
(f) Scaling of the entanglement entropy with respect to the correlation lengths at the critical point, yielding a central charge $c=1$.}
\label{tx0.7fig}
\end{figure}

When $t_2=0$, the model in Eq.(\ref{H_spin}) represents the quantum spin-1/2 antiferromagnetic XXZ chain, which undergoes a BKT transition from the XY phase ($V/t_1 < 2$) to the AFM (density wave (DW) with all particles located in one of the legs in the bosonic language) phase ($V/t_1 > 2$) \cite{sun2015fidelity}. 
Our numerical results obtained from the fidelity susceptibility confirm that the critical value is located at $V/t_1 = 2$, as illustrated in Fig.\ref{phasefig}. 
We then investigate whether this phase transition persists under the next-nearest-neighbor hopping using the DMRG. 
The fidelity susceptibility is computed as a function of $V/t_1$ at $t_2=0.1$ for large open chains. 
The numerical results indicate that the peak of the fidelity susceptibility gradually shifts towards the left, and its magnitude slowly increases with increasing system size, as shown in Fig.\ref{tx0.1fig}(a). 
Finite-size scaling analysis, presented in Fig.\ref{tx0.1fig}(c) and (d), demonstrates that the fidelity susceptibility continues to adhere to the scaling laws proposed for BKT transitions \cite{sun2015fidelity}, as outlined in Eq.(\ref{FS}) and Eq.(\ref{FS2}). 
Moreover, it is observed the staggered magnetization in the $z$-direction,
\begin{align}
O_{\text{AFM}} &= \frac{1}{L} \sum_{r} (-1)^{r} \langle \sigma_r^z \rangle, \label{OP:AFM}
\end{align}
begin to increase from zero as the interaction $V$ increases [cf. Fig.\ref{tx0.1fig}(b)], indicating the existence of the AFM phase in the $z$-direction for $V > V_c$.

\section{Deconfined quantum critical point}  
We have demonstrated that the BKT transitions between the XY phase and VBS state, as well as the BKT transition between the XY phase and AFM phase, persist even in the presence of small next-nearest-neighbor hopping ($t_2/t_1$) and interaction ($V/t_1$). Now, we turn to explore the transitions in the regime characterized by large values of $t_2/t_1$ and $V/t_1$. 
Interestingly, we find unexpectedly that the system undergoes a phase transition from the AFM phase to the VBS state as $t_2/t_1$ or $V/t_1$ increase further.

As an example, we compute the fidelity susceptibility and the order parameters at $t_2/t_1=0.7$ [cf. Fig.\ref{tx0.7fig}]. It is observed that the system resides in the VBS phase for $0<V/t_1<2.12$, characterized by a non-zero $O_{\text{VBS}}$ and vanishing $O_{\text{AFM}}$. Conversely, in the regime $2.12<V/t_1<3$, the reverse occurs, indicating that the system is in the AFM phase [cf. Fig.\ref{tx0.7fig}(b)]. 
Notably, both order parameters, $O_{\text{AFM}}$ and $O_{\text{VBS}}$, seem to smoothly vanish at this same point, suggesting an unconventional nature for this quantum phase transition. 
This transition is argued to be a second-order phase transition with the correlation length critical exponent $\nu=1.11$ (see Appendix \ref{AppC} for additional critical exponents), as demonstrated in Fig.\ref{tx0.7fig}(a), (c), and (d) from the finite-size scaling of the fidelity susceptibility \cite{gu2010fidelity,sun2017fidelity,zhu2018fidelity},
\begin{align}
   \chi_L^{m} \propto L^{2/\nu-1},
   \label{FS:2nd}
\end{align}
where $\chi_L^{m}$ represents the fidelity susceptibility per site at the peak position $\lambda=\lambda_{m}$, which converges towards the critical point $\lambda_c$ as $L$ tends to infinity. 
As the AFM phase breaks the $\mathbb{Z}_2$ spin-flip symmetry and the VBS phase breaks translation symmetry, the direct continuous second-order phase transition occurring between the AFM phase and the VBS state indicates that the phase transition is a DQCP beyond the LGW theory. 
The continuous phase transition is further confirmed by the divergence of the correlation length $\xi$ with the increasing number of states $M$ [cf. Fig.\ref{tx0.7fig}(e)], as computed using TenPy \cite{tenpy2024}. Additionally, the central charge $c=1$ is derived from the scaling of the entanglement entropy $S$ [cf. Fig.\ref{tx0.7fig}(f)], using the relation $S=\frac{c}{6} \log(\xi)$ \cite{huang2019emergent,lee2023landau,romen2024deconfined,baldelli2024frustrated,Roberts2019}.
We note that the results obtained from iTEBD method are consistent with those from finite-size scaling (see also Appendices \ref{AppD} and \ref{AppE} for details).

Interestingly, the DQCP arises from the merging of two BKT transitions \cite{lee2023landau,romen2024deconfined,baldelli2024frustrated}. These transitions can be understood through the concept of domain walls, which are topological defects in one-dimensional systems. 
Beginning from the VBS phase (i.e at $t_2 / t_1=0.7$ and $V=0$), increasing the interaction strength $V/t_1$ induces the formation of AFM domain walls. Domain walls in VBS bind the AFM phase (and vice versa), leading to the destruction of the VBS phase and the emergence of the AFM phase as $V/t_1$ increases further. Consequently, a direct transition between the VBS phase to the AFM phase occurs under conditions of large $t_2/t_1$ and $V/t_1$, indicating the presence of a DQCP. This DQCP is linked to the two BKT transitions through a multicritical point characterized by an emergent high symmetry.

We note that our finding is analogous to the phenomena described in frustrated two dimentional models \cite{liu2022emergence,liu2024emergent}, where a gapless quantum spin liquid (QSL) phase gradually develops from the AFM-VBS transition, leading to two phase transitions between gapped and gapless phases (AFM-QSL transition and QSL-VBS transitions). Consequently, a DQCP emerging from two BKT transitions may be a universal phenomenon, offering crucial insights into the understanding of both BKT transitions and DQCPs.

\section{Conclusion} 
In summary, we employed the DMRG and iTEBD methods to investigate novel physical phenomena in the frustrated dipolar Bose-Hubbard model and to provide a comprehensive phase diagram of the model through the exploration of fidelity susceptibilities and order parameters. We identified two gapped phases, the AFM phase and the VBS phase, along with one gapless XY phase, each exhibiting two BKT transitions and a DQCP. Remarkably, we found that the DQCP arises from the fusion of two BKT transitions in this model. Moreover, we propose that this model could be realized using dipolar bosons.

\begin{acknowledgments}
G.S. thanks Luis Santos for the valuable comments.
G.S. is appreciative of support from the NSFC under the Grants No. 11704186, "the Fundamental Research Funds for the Central Universities, NO. NS2023055". 
W.-L.Y is supported by the NSFC under Grant No. 12174194, Opening Fund of the Key Laboratory of Aerospace Information Materials and Physics (Nanjing University of Aeronautics and Astronautics), MIIT, Top-notch Academic Programs Project of Jiangsu Higher Education Institutions (TAPP), and stable supports for basic institute research under Grant No. 190101. 
Y.-N.W and W.-Y.Z are supported by the Postgraduate Research \& Practice Innovation Program of Jiangsu Province under Nos. KYCX24\_0526 and KYCX23\_0347, respectively.
S.-P. K is appreciative of support from the NSFC under the Grant Nos. 11974053 and 12174030, and the National Key R\&D Program of China under Grant No. 2023YFA1406704.
This work is partially supported by the High Performance Computing Platform of Nanjing University of Aeronautics and Astronautics.
\end{acknowledgments}

\bibliography{ref}

\begin{widetext}
\appendix
\setcounter{figure}{0}
\renewcommand{\thefigure}{A\arabic{figure}}
\section{Non-interacting energy spectrum}
\label{AppA}
We first discuss the case of free bosons ($V=0$ and $U=0$), which can be simplified as follows:
\begin{equation}
    H=\sum_r t_1(b_r^{\dagger}b_{r+1}+ \text{h.c.})+t_2(b_r^{\dagger}b_{r+2}+ \text{h.c.})
\label{SA:eq2}
\end{equation}
 Using the Fourier transformation,
 \begin{equation}
 \begin{split}
     b_r=\frac{1}{\sqrt N}\sum e^{ikj}b_k, \\
     b_r^{\dagger}=\frac{1}{\sqrt N}\sum e^{-ik'j}b_{k'}^{\dagger},
\end{split}
\end{equation}
the model described in Eq.(\ref{SA:eq2}) can be transformed into momentum space,
\begin{equation}
\begin{split}
H &=t_1\sum_{kk'}\delta_{kk'}e^{ik}b_{k'}^{\dagger}b_k + t_1\sum_{kk'}\delta_{kk'}e^{-ik'}b_{k'}^{\dagger}b_k + t_2\sum_{kk'}\delta_{kk'}e^{2ik}b_{k'}^{\dagger}b_k + t_2 \sum_{kk'}\delta_{kk'}e^{-2ik'}b_{k'}^{\dagger}b_k \\
   &=\sum_{k} \left[ t_1(e^{ik}+e^{-ik})+t_2(e^{2ik}+e^{-2ik}) \right] b_{k}^{\dagger}b_k\\
   &=\sum_{k} \left( 2t_1\cos{k}+2t_2\cos{2k} \right) b_{k}^{\dagger}b_k \\
   &=\sum_{k} \epsilon(k) b_{k}^{\dagger}b_k \\
\end{split}
\end{equation}

\begin{figure}[h]
\includegraphics[width=8cm]{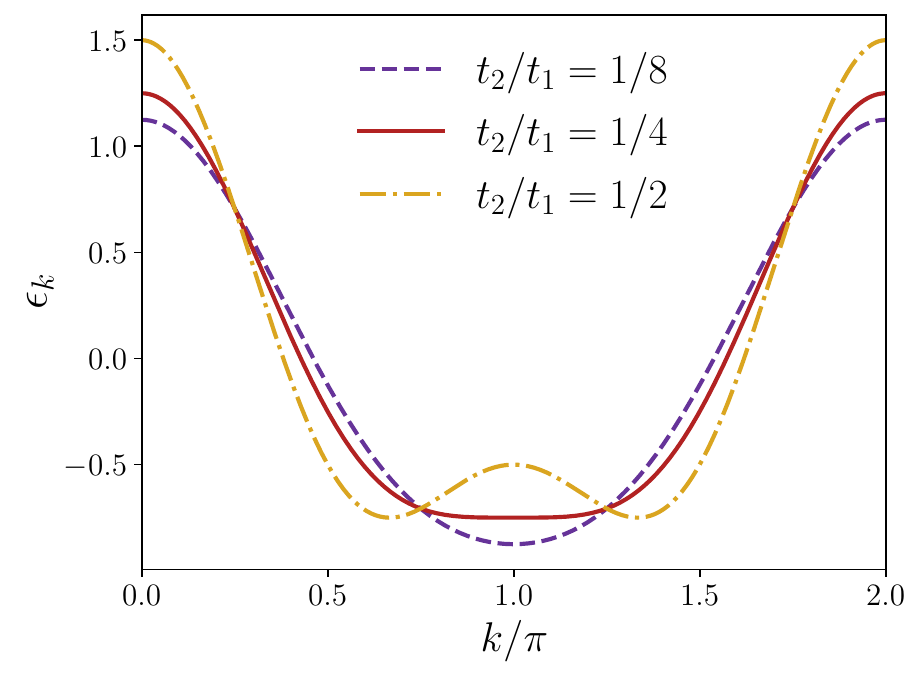} 
\centering
\caption{ The energy spectrum of free bosons with different $t_2/t_1$.}
\label{SA:E0fig}
\end{figure}

The Hamiltonian is diagonalized, with the spectrum,
\begin{equation}
\epsilon(k)=2 t_1 (\cos{k}+j\cos{2k}),
\end{equation}
Here, $j={t_2}/{t_1}$. 
Using trigonometric relationships, $\cos{2k}=2\cos^{2}k-1$, the spectrum can be transformed as:
\begin{equation}
\epsilon(k)=2 t_1 (2j\cos^{2}k+\cos{k}-j).
\end{equation}
The minimum of $\epsilon(k)$ is obtained by differentiating the derivative with respect to $k$, which gives:
\begin{equation}
    \frac{d}{dk}\epsilon(k)=-2 t_1(4j\cos{k}\sin{k}+\sin{k}) = 0.
\end{equation}
The equations $\sin{k}=0$ and $1+4j\cos{k}=0$ offer two solutions, thereby affecting the spectrum $\epsilon(k)$ depending on the frustration parameter $j$ \cite{Greschner2013ultracold}. Specifically:
When $j < \frac{1}{4}$, the dispersion $\epsilon(k)$ displays a single minimum at $k=\pi$ because $\sin{k}=0$.
Conversely, when $j > \frac{1}{4}$, the spectrum $\epsilon(k)$ exhibits two distinct minima located at $k = \pm \arccos\left(-\frac{1}{4j}\right)$, arising from the condition $1+4j\cos{k}=0$. 
The critical value $j = \frac{1}{4}$ marks a special point known as the Lifshitz point. The system exhibits either a superfluid phase when $j < \frac{1}{4}$ or a chiral superfluid phase when $j > \frac{1}{4}$.

Next, we consider bosons in the hard-core limit. The Hamiltonian subject the hard-core constraint $(U\rightarrow\infty)$ is,
\begin{equation}
 H=\sum_r t_1(b_r^{\dagger}b_{r+1}+\text{h.c})+t_2(b_r^{\dagger}b_{r+2}+ \text{h.c.})
\label{fermion}
\end{equation}
with $n_r=b_r^{\dagger}b_{r}=\{ 0, 1 \}$ represent particle number operators with a hard-core constraint at site $r$.
In the hard-core limit with infinite interaction, only one boson can occupy a single lattice site, behaving similarly to spinless fermions. Therefore, the superfluid phase and the chiral superfluid phase observed in free bosons are expected to transform into a gapless phase and a phase characterized by symmetry breaking.

\setcounter{figure}{0}
\renewcommand{\thefigure}{B\arabic{figure}}
\section{Level crossing for the BKT transition}
\label{AppB}

\begin{figure}[h]
\includegraphics[width=9cm]{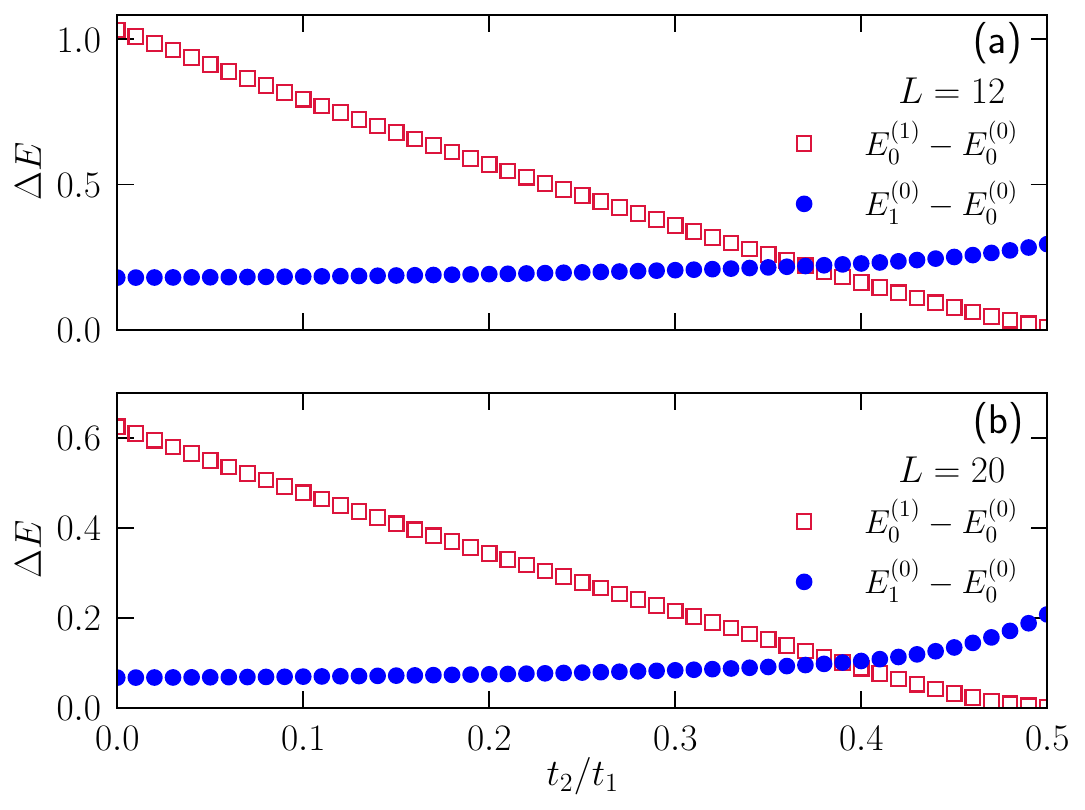} 
\centering
\caption{The energy difference of the FDBH model with respect to the next-nearest-neighbor hopping. The red squares represent the energy difference between the ground state and the first excited state with $S_{\text{total}}^z=0$. The blue circles represent the energy difference between the ground state of $S_{\text{total}}^z=1$ and $S_{\text{total}}^z=0$. For $L=12$, the crossing point occurs at $t_{2c}$=0.37. For $L=20$, the crossing point locates at $t_{2c}=0.39$. }
\label{SB:delta_E}
\end{figure}

To further pinpoint the critical point of the system, we calculated the distributions of several lowest energy levels by imposing U(1) symmetry, specifically conserving total spin along the $z$-axis (equivalent to conserving total particle numbers in bosonic language), 
\begin{equation}
S_{\text{total}}^z=\sum_{r}{S_r^z}.
\end{equation}
When the number of particles in the system is a multiple of four, the ground state is a non-degenerate singlet state \cite{nomura1994critical}. In the XY phase, the first excited state exhibits $S_{\text{total}}^z=1$, whereas in the VBS phase, the first excited state exhibits $S_{\text{\text{total}}}^z=0$.
Therefore, the XY-VBS phase transition can be identified by the level crossing between the doublet and the singlet of the first excited states.

To identify the phase transition via level crossings, we investigated the energy gap $\Delta E = E_m^{(n)} - E_{m^\prime}^{(n^\prime)}$ of the FDBH model using exact diagonalization with lattice sizes $L=12$ and $L=20$ at $V=0.1$ for $0<t_2/t_1<0.5$ as illustrated in Fig.\ref{SB:delta_E}. Here, $E_m^{(n)}$ denotes energy levels where $m=0,\pm1$ represents the total spin along the $z$-direction, and $n$ denotes the energy level (e.g., ground state, first excited state).
The red squares depict the energy difference between the singlet ground state and the first excited state with $S_{\text{total}}^z=0$, 
which is non-degenerate except at the Majumdar-Ghosh point $t_2=0.5$.
The blue circles correspond to the energy difference between ground states with $S_{\text{total}}^z=1$ and $S_{\text{total}}^z=0$. The critical point is identified where these two lines cross. For $L=12$, the XY phase to VBS phase transition occurs at $t_{2c}=0.37$, while for $L=20$, it shifts slightly to $t_{2c}=0.39$. This shift to higher $t_{2c}$ values with increasing lattice size indicates a pronounced finite-size effect.
In the hard-core boson limit, spin conservation is related to particle number conservation through the equation,
\begin{equation}
 S_{\text{total}}^z = L/2 - N_{\text{total}},   
\end{equation}
where $L$ is the total number of lattice sites and $N_{\text{total}}$ represents the total number of particles in the system. This mapping allows us to interpret spin states in terms of particle configurations. For instance, when $S_{\text{total}}^z=0$, the system corresponds to a half-filled particle case. By varying the particle number within a specific subspace, one can identify excited states.

\setcounter{figure}{0}
\renewcommand{\thefigure}{C\arabic{figure}}
\section{Correlation length critical exponent}
\label{AppC}

\begin{figure}[h]
\includegraphics[width=8cm]{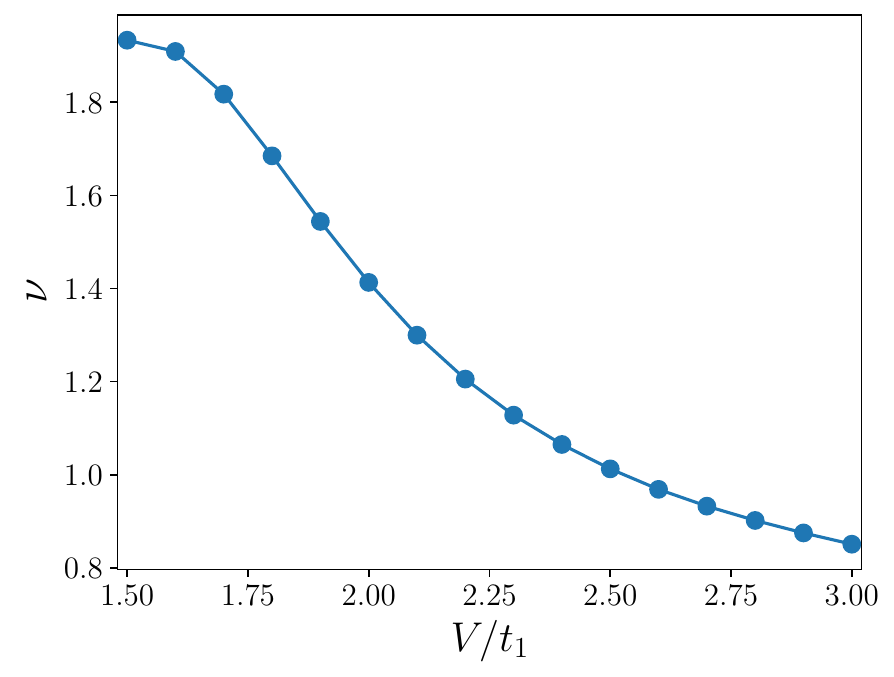} 
\centering
\caption{The critical exponents of the correlation length for VBS-AFM transitions.}
\label{mufig}
\end{figure}

Quantum many-body systems demonstrate finite-size scaling and universal behaviors near quantum critical points, allowing for the extraction of critical exponents. The fidelity susceptibility near the critical point becomes more pronounced as the system size increases, highlighting significant finite-size scaling effects at criticality. It is well-established that the fidelity susceptibility exhibits finite-size scaling near the critical point,
\begin{equation}
    \chi_{F}(\lambda_m)\sim L^{2 / \nu},
    \label{nu}
\end{equation}
for second-order phase transitions. Here, $\nu$ denotes the correlation length critical exponent, and $\lambda_m$ represents the peak position of the fidelity susceptibility for a system of size $L$. 
The correlation length critical exponent is determined by fitting the logarithm of both sides of Eq. (\ref{nu}):
\begin{equation}
    \ln \chi_{F}(\lambda_m)\propto 2 / \nu \ln L.
    \label{mulnL}
\end{equation}
Here, $ \ln \chi_{F}(\lambda_m)$ corresponds to the natural logarithm of the fidelity susceptibility evaluated at the peak position $\lambda_m$, and 
$L$ denotes the system size.

In the vicinity of the phase transition, we keep the interaction strength fixed and compute the ground state fidelity susceptibility of the system across various sizes (including $L=128, 144, 192, 240$, and $288$)  as a function of the next-nearest-neighbor hopping $t_2/t_1$. By fitting the maximum fidelity against system size using the logarithmic relationship in Eq. (\ref{mulnL}), we extract the slope of the fit, thereby determining the correlation length critical exponent, as illustrated in Fig. \ref{mufig}. From the figure, it is observed that the critical exponents decrease gradually with increasing interaction strength. This suggests that the system exhibits varying universality classes for different quantum critical points.

\setcounter{figure}{0}
\renewcommand{\thefigure}{D\arabic{figure}}
\section{Correlation functions}
\label{AppD}

\begin{figure}[h]
\includegraphics[width=16.0cm]{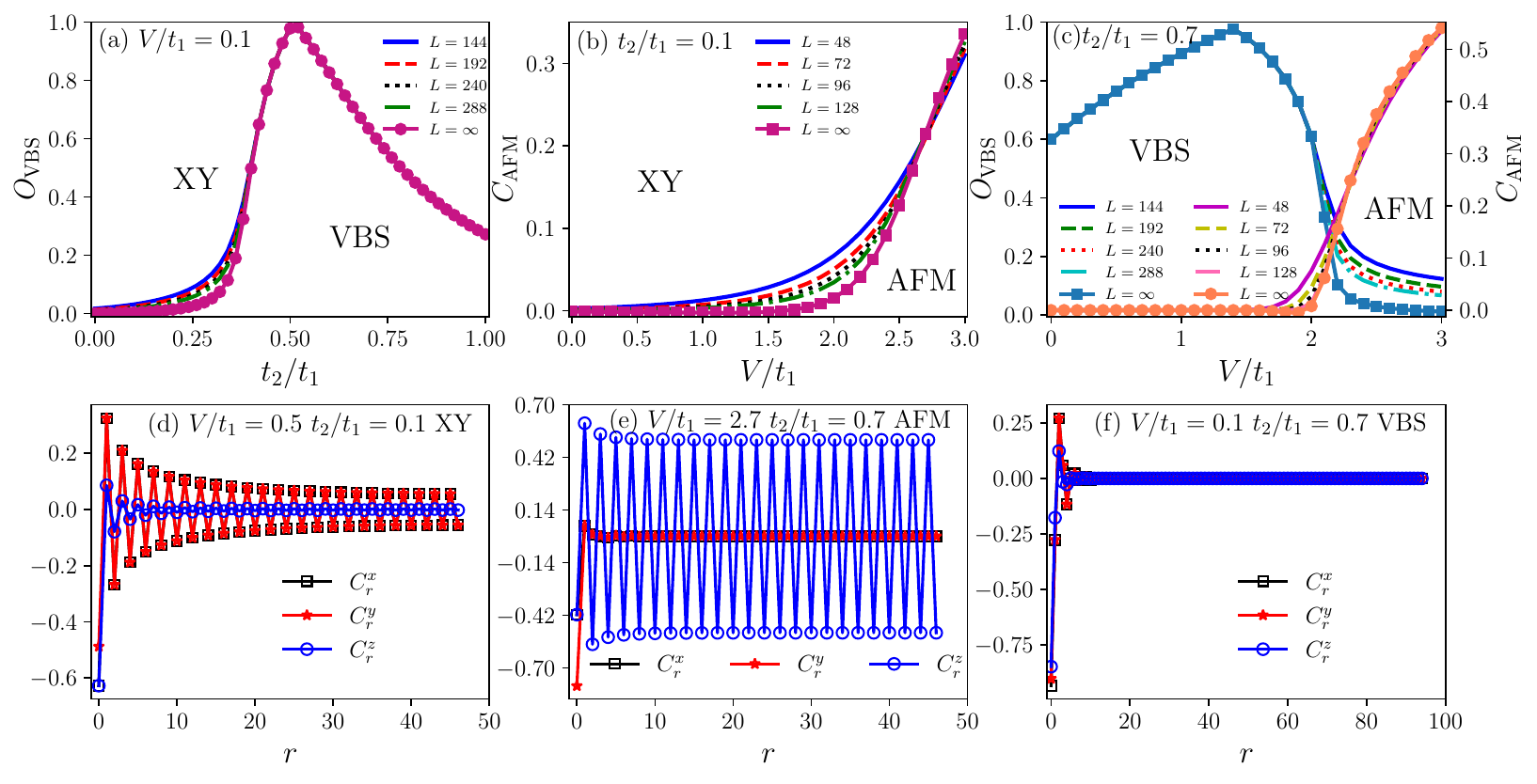}
\centering
\caption{Order parameters and correlation functions.
(a) XY to VBS phase transition at $V/t_1=0.1$ as the function of $t_2/t_1$,
(b) XY to AFM phase transition at $t_2/t_1=0.1$ as the function of $V/t_1$,
(c) VBS to AFM phase transition at $t_2/t_1=0.7$ as the function of $V/t_1$,
(d) Correlation functions for XY phase at $t_2/t_1=0.1, V/t_1=0.5$ for $L=96$ in PBCs,
(e) Correlation functions for AFM phase at $t_2/t_1=0.7,V/t_1=2.7$ for $L=96$ in PBCs,
(f) Correlation functions for VBS phase with $t_2/t_1=0.7, V/t_1=0.1$ for $L=192$ in OBCs.}
\label{CorrelationFunction}
\end{figure}

In this section, we calculated the long-range correlation functions defined as,
\begin{equation}
C^{\alpha}_{r} = \langle\sigma^\alpha_{1} \sigma^\alpha_{r+1}\rangle,
\end{equation}
with $\alpha=x,y,z$ for each phase in order to demonstrate the characteristics of each phase. Additionally, we compute the order parameters for finite size systems.
The order parameters $O_{\text{VBS}}$ for the VBS phase is computed from Eq.(\ref{OP:VBS}),
while the order parameters $O_{\text{AFM}}$ for the AFM phase can be derived in terms of correlation functions as \cite{rossini2012phase,mbeng2024quantum},
\begin{equation}
C_{\text{AFM}} = \lim_{r \rightarrow \infty} C^{z}_{r} \rightarrow O_{\text{AFM}}^{2}.
\end{equation}
In the following, I will present the order parameters $C_{\text{AFM}}$ based on above correlation functions. 
In our calculations, $C_{\text{AFM}}$ is computed using $r=L/2$ for system sizes up to $L=128$ in periodic boundary conditions (PBCs), while the order parameters $O_{\text{VBS}}$ is obtained for system sizes up to $L=288$ under OBCs. 
The results for $C_{\text{AFM}}$ and $O_{\text{VBS}}$ are presented in Fig.\ref{CorrelationFunction}(a-c), where it can be observed that they are consistent with the iTEBD results [c.f. Fig.\ref{V0.1fig}, Fig.\ref{tx0.1fig} and Fig.\ref{tx0.7fig}].
The nature of the XY, AFM and VBS phases is further confirmed by the long-range correlation functions, as demonstrated in Fig.\ref{CorrelationFunction}(d-f).

\setcounter{figure}{0}
\renewcommand{\thefigure}{E\arabic{figure}}
\section{DMRG simulations}
\label{AppE}

In this section, we discuss the convergence of the ground state energy of the system. 
During the DMRG simulations, we gradually increase the number of states to a maximum of $M=500$ for different lattice sizes, as well as the number of DMRG sweeps to a maximum of 40 until the system converged. We benchmark our results by comparing the energy differences of the system at $t_2/t_1=0.7$ for various states as shown in Fig.\ref{DMRGSim}. 
We find that under OBCs, the energy difference between $M=400$ and $M=500$ states for $L=288$ is very small, reaching $10^{-9}$, while under PBCs, the energy difference between $M=400$ and $M=500$ states for $L=96$ is also very small, reaching $10^{-5}$. 
The results from the finite-size DMRG simulations are in good agreement with the iTEBD results.

\begin{figure}[h]
\includegraphics[width=9.0cm]{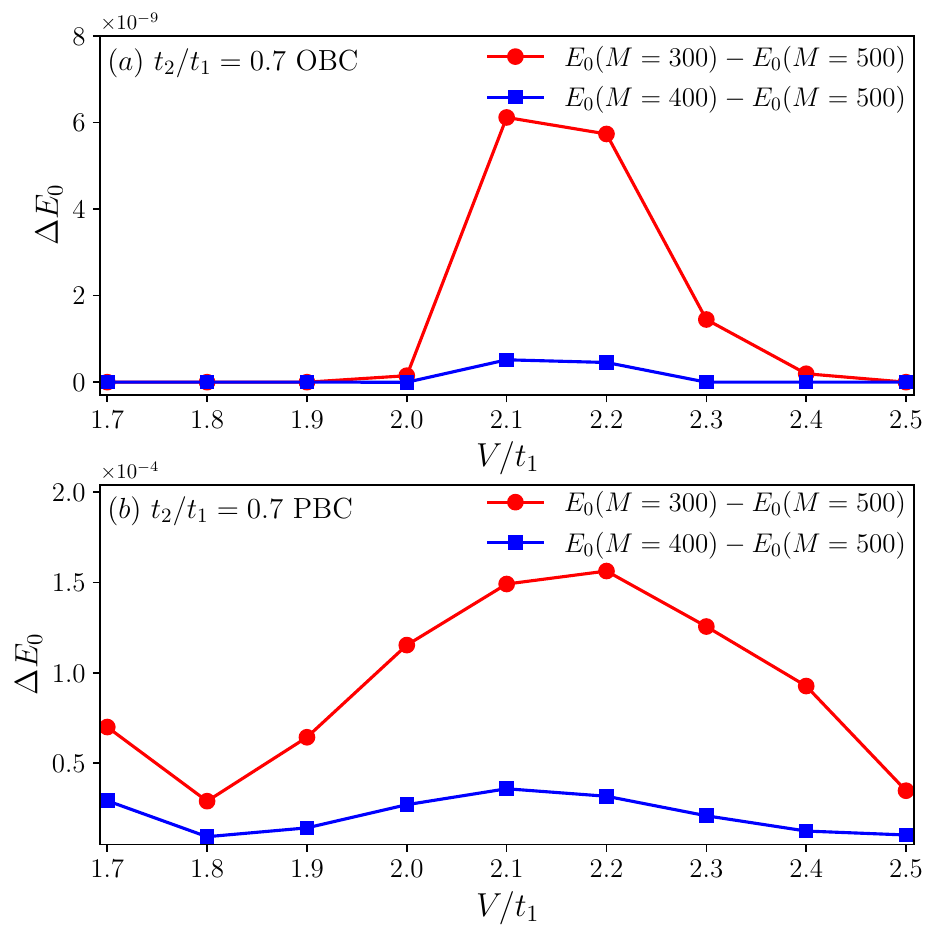}
\centering
\caption{The difference in ground state energy. (a) the energy difference $\Delta E_{0}$ for $L=288$ at $t_2/t_1=0.7$ under OBCs, (b) the energy difference $\Delta E_{0}$ for $L=96$ at $t_2/t_1=0.7$ under PBCs.}
\label{DMRGSim}
\end{figure}

\end{widetext}

\end{document}